\documentclass[12pt]{article}
\usepackage[dvips]{graphicx,color}
\begin{document}
\title{%
$\gamma$-ray emission from Outer-Gap of pulsar magnetosphere}

\author{Junpei Takata and Shinpei Shibata\\
{\it University of Yamagata, 1-4-12, Yamagata, Yamagata, 990-8560, Japan}\\
Kouichi Hirotani\\
{\it Code 661, Lab. for High-energy Astrophys.,}\\ 
{\it NASA/GSFC, 
Greenbelt, MD 20771}}

\maketitle
\section*{Abstract}
We develop a model for $\gamma$-ray emission from the outer magnetosphere of 
pulsars (the outer-gap model). The charge depletion causes 
a large electric field which 
accelerates electrons and positrons. We solve the
 electric field with radiation 
and pair creation processes self-consistently, and calculate curvature 
spectrum and Inverse-Compton (IC) spectrum. We apply this 
theory to PSR B0833-45 (Vela) and B1706-44 for which their surface magnetic 
fields, observed thermal X-rays are similar to each other. We 
find that each observed cut-off 
energies of the $\gamma$-rays are well explained. 
By inclusion of emission outside the gap, 
the spectrum is in better
 agreement with the observations than the spectrum arising only from the 
inside of the gap.  
The expected TeV fluxes are much smaller than 
that observed by CANGAROO group in the direction 
of B1706-44.
\section{Introduction}
The Compton Gamma Ray Observatory has detected seven $\gamma$-ray pulsars. 
The observed light curves and energy spectra have been used to discriminate 
possible radiation models. 
Moreover, the next-generation $\gamma$-ray space telescopes 
and the ground based Cherenkov telescopes 
will further constrain to the models. 

The pulsed $\gamma$-rays imply
 that the electrons and positrons
 are accelerated up to about 10 TeV in the magnetosphere. 
For the outer-gap model (Chen, Ho \& Ruderman, 1986),  
Hirotani \& Shibata (1999, HS) solved the accelerating electric 
field with curvature 
radiation and pair creation processes self-consistently. 
They showed that 
the electric field along the magnetic field does not extend to the 
light cylinder, and calculated 
curvature spectra can explain the EGRET 
observations in the GeV bands. However, the gap emission could not explain 
the observations in the MeV band. 
Hence, 
to improve the HS model, we take account of the radiation from the outside
 of the gap and compare the corrected spectrum with the observations.
\section{One dimensional model}
In this section, we introduce the HS model and represent 
the Poisson eq. which describes the 
accelerating electric field, 
the continuity eqs. for particles and $\gamma$-rays 
, following HS. 
\subsection{Basic equations}
We deal with the structure along the magnetic field lines.
 If the gap width ($W$) along the magnetic field is much less than 
the light radius ($\varpi_{lc}$), 
we can approximate the magnetic field lines as straight lines 
in the gap, and the electric field structures can be treated 
as one-dimensional, where the arc length from
 the surface along the last-closed line is denoted by $s$ : 
we can write down the Poisson equation as
\begin{equation}
\frac{dE_{||}}{us}=4\pi e\left(N_+-N_--
\frac{\rho_{GJ}}{e}\right),
\label{poission}
\end{equation}
where $E_{||}$ is the electric field along the magnetic field, $N_+$ ($N_-$) 
is the electron (positron) number density, 
and $\rho_{GJ}$ is the Goldreich-Julian (GJ) 
charge density. Above equation describes that the charge depletion relative to 
$\rho_{GJ}$ causes $E_{||}$.

By taking account of the electron-positron pair creation process, the 
 continuity equations for particles and $\gamma$-rays are 
\begin{equation}
\pm B\frac{d}{ds}\left(\frac{N_{\pm}}{B}\right)
=\frac{1}{c\cos\Psi}\int_0^{\infty}
d\epsilon_{\gamma}[\eta_pG_{+}+\eta_-G_-],
\label{con1}
\end{equation}
\begin{equation}
\pm B\frac{d}{ds}\left(\frac{G_{\pm}}{B}\right)
=\frac{-\eta_{p\pm}G_{\pm}+\eta_cN_{\pm}}{c\cos\Psi},
\label{con2}
\end{equation}
where $G_+$ ($G_-$) is the distribution function of outward (inward) 
propagating $\gamma$-rays, 
$\epsilon$ refers to the photon energy in units of the 
electron's rest mass energy, 
$B$ is the magnetic field strength, $\eta_{p\pm}$ is the pair-creation rate,
$\eta_c$ is the emissivity of curvature radiation,  
and $\Psi$ is the angle between 
the particle's motion and the meridional plane. 
We describe $G_{\pm}$ in several 
energy bins and represent then as $G_{\pm}^i$ $(i=1,2,...)$.
The particle's Lorentz factor in the gap is obtained 
by assuming that particle's motion immediately saturated at the balance 
 between the electric and the radiation reaction forces, i.e.,
\begin{equation}
\Gamma_{sat}=\left(\frac{3R_c^2}{2e}E_{||}\right)^{1/4},
\label{lorent}
\end{equation} 
where $R_c$ is the curvature radius of dipole magnetic field lines.

We impose some boundary conditions. 
The inner ($s_1$)
and outer ($s_2$) boundaries are defined so that $E_{||}$ vanishes, i.e., 
$E_{||}(s_1)=E_{||}(s_2)=0$. 
We assume that the $\gamma$-rays do not come into the gap 
through the boundaries, 
$G_{+}^{i}(s_1)=G_{-}^{i}(s_2)=0$ $(i=1,2,....m)$. We allow the 
particles to come into the gap. The particle flux is given by the 
non-dimensional parameters $j_1$ and $j_2$ as follows:
\begin{equation}
\frac{N_{+}(s_1)}{\Omega B(s_1)/2\pi ce}=j_1
,
\frac{N_{-}(s_2)}{\Omega B(s_2)/2\pi ce}=j_2.
\label{cond}
\end{equation}
The particle continuity equation (\ref{con1}) yields  
\begin{equation}
\frac{N_{+}(s)}{\Omega B(s)/2\pi ce}+\frac{N_{-}(s)}{\Omega B(s)/2\pi ce}
=\textrm{const along s} =j_{tot}.
\label{conti}
\end{equation}
From eqs. (\ref{cond}) and (\ref{conti}), the current carriers 
created in the 
gap per unit flux tube is $j_{gap}=j_{tot}-j_1-j_2$.
 The total current should be determined by the global condition 
which includes pulsar wind.
Therefore, we use $(j_{tot},j_1,j_2)$ as free model parameters.

\subsection{X-ray \& infrared (IR) field}
Because the $\gamma\gamma$ 
pair-creation process is important in the outer magnetosphere, 
we need the X-rays for the target-photons in our case. 
In the present paper, we use the observed black body radiation 
from the pulsar surface for the X-rays (Table 1). 
We also need the IR field to calculate IC flux.
The IR field is inferred from optical and radio observations for Vela,  
X-ray and radio observations for 
B1706-44 with a single power-low, because there are 
no available IR observations.
\begin{table}[t]
\caption{Observed Parameters}
\begin{center}
\begin{tabular}{cccccc}
\hline\hline
pulsar & distance & $\Omega$ & $B_{12}$ & $kT_s$ & Ref.\\
 & kpc & rad/s & $10^{12}G$ & eV &\\
\hline
Vela & 0.5 & 70.6 & 3.4 & 150 & \"{O}gelman et. al.\\
B1706-44 & 1.8(DM)/2.5(HI) & 61.6 & 3.1 &143 &Gotthelf et.al.\\

\hline
\end{tabular}

DM : Dispersion Measure , HI : HI absorption
\end{center}
\end{table}

\section{Radiation from outside of the gap}
Near the boundaries, the real Lorentz factor of the accelerated 
particles must be 
lager than $\Gamma_{sat}$ given
 by eq.(\ref{lorent}), because the particle's cooling time for 
the radiation becomes larger than the crossing time for the gap.
So, the particles come out from the gap with $\Gamma_{out}\sim$ 
$a$ $few$ $times$ $10^7$ and emit $\gamma$-rays outside of the gap. 
Since the typical damping length for the curvature radiation is 
\begin{equation}
l_{dam}=\frac{3}{2}\frac{m_ec^2}{e^2}\frac{R_c^2}{\Gamma^3}
=0.4\varpi_{lc}\left(\frac{\Omega}{100\textrm{rad}\mathrm{s^{-1}}}
\right)^{-1}
\left(\frac{\Gamma}{10^7}\right)^{-3}\left(\frac{R_c}{0.5\varpi_{lc}}\right)^2,
\end{equation}
 the $\gamma$-ray radiation from the outside of the gap are 
also important unless $W\sim\varpi_{lc}$.

We apply the our model to Vela and B1706-44 
with the observed parameters in Table 1.
We adopt $(j_{tot},j_1,j_2)=(0.201,0.191,0.001)$ and $a_{inc}=45^{\circ}$ for 
the angle between axes of rotation and magnetization for both pulsars.
\section{Result}

\subsection{Electric field structure}
The calculated $E_{||}$ for both pulsars are shown in Fig.1. 
The gap width $W$ is shorter than 
$\varpi_{lc}$. $W$ is characterized by the pair-creation mean free path, 
which is given approximately by 
$W\underline{\propto} c/(\int\eta_{p-}\eta_cd\epsilon)^{1/2}$ by using
 the fact $\eta_{p+}\ll\eta_{p-}$, due to  the difference 
in the collision angle between $\gamma$-ray and X-ray. For Vela and B1706-44, 
one finds the mean free path to be shorter than $\varpi_{lc}$.

   We find that Vela has a nearer distance from the 
surface to the gap and larger calculated $E_{||}$ than B1706-44
 when we adopt the same $(j_{tot},j_1,j_2)$ and $a_{inc}$.
 This is because Vela has the shorter rotation period and larger GJ density 
in the gap than B1706-44.  
The dependence of the assumed distance from the 
earth to B1706-44 is also shown in Fig.1. In general, 
if we adopt a nearer distance to the pulsar, 
the electric field becomes large, because the decrease in the estimated X-ray 
luminosity from the observations extends the gap width.
 
\begin{figure}[t]
 \begin{center}
\includegraphics[height=13pc]{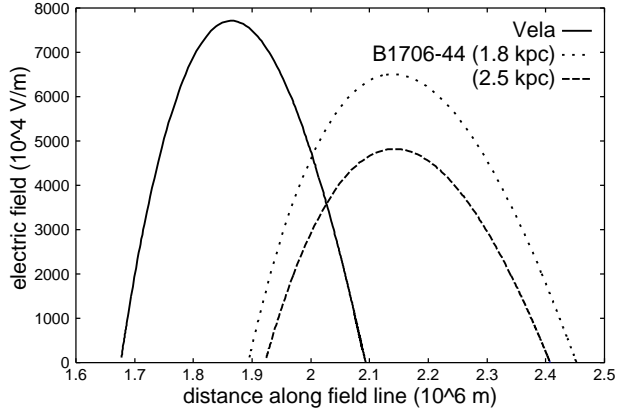}
\end{center}
\caption{The accelerating field for Vela (solid-line) and B1706-44 
(dotted-line for dis=1.8kpc and dashed-line for dis=2.5kpc). 
$(j_{tot},j_1,j_2)=(0.201,0.191,0.001)$ and $a_{inc}$=45 deg.}
\end{figure}
\begin{figure}[h]
  \begin{center}
    \includegraphics[width=16pc,height=13pc]{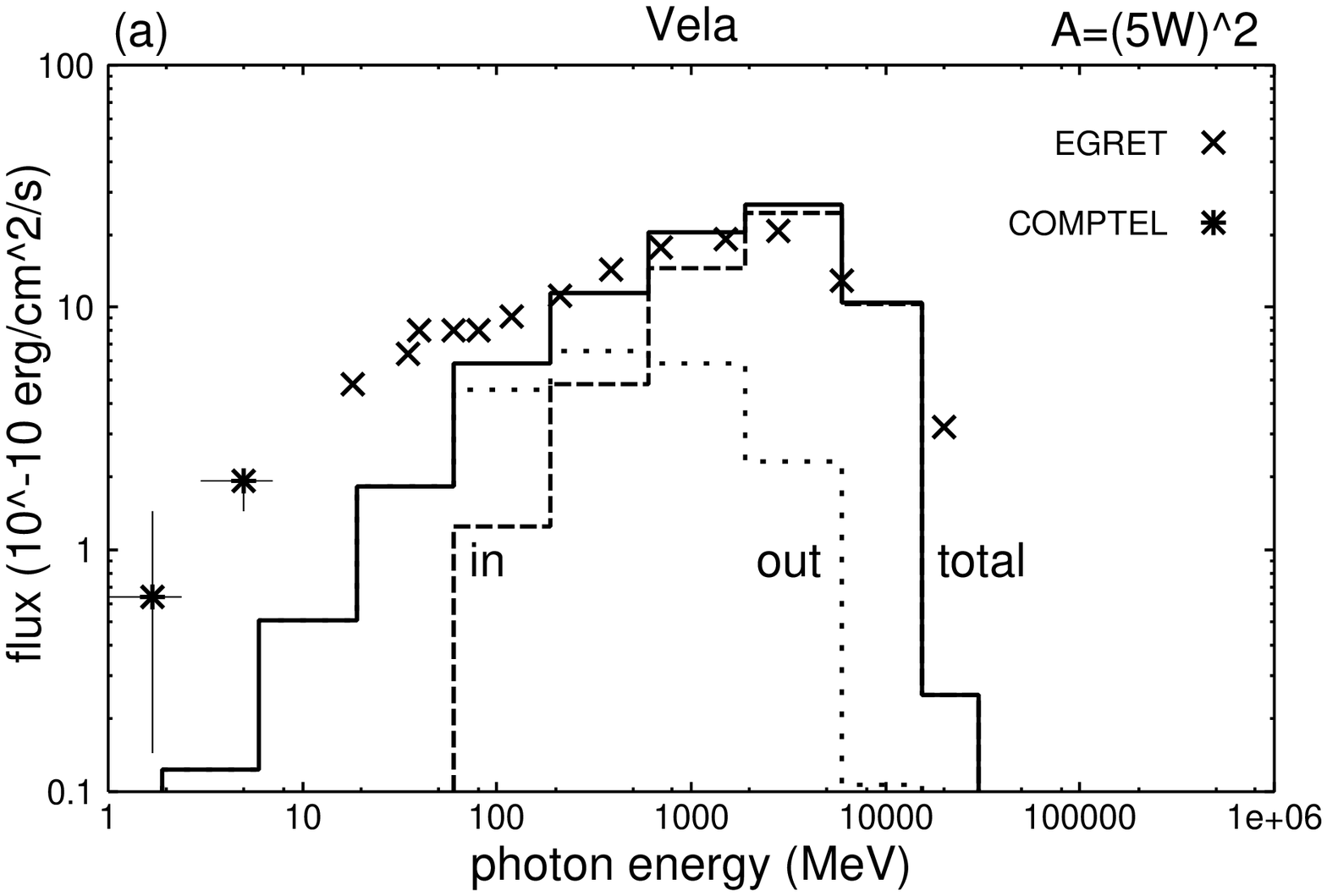}
 \includegraphics[width=16pc,height=13pc]{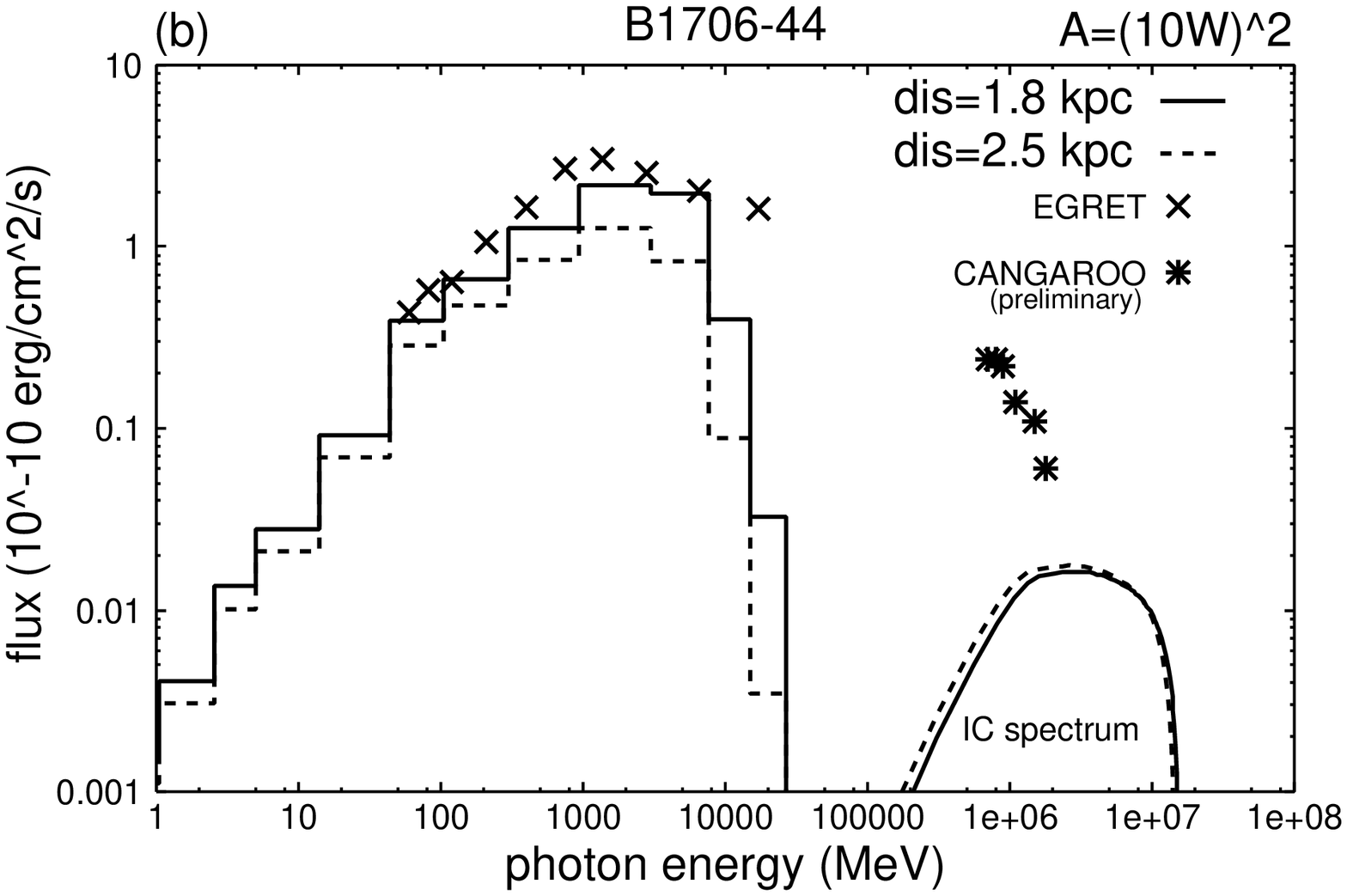}
  
\caption{(a):$\gamma$-ray spectrum for Vela. The total spectrum (solid-line)
 includes the radiation from the outside of the gap (dots-line) as well as gap emission (dashed-line). (b):Total $\gamma$-ray spectrum for B1706-44. The dependence of distance is shown. The IC spectrum is also shown in figure.}   
\end{center}
\end{figure}
\subsection{Gamma-ray spectra}
The calculated spectrum of outward 
propagating $\gamma$-rays radiated in the gap for Vela is 
shown in Fig.2(a) as dashed-line. The spectral cut-off 
around GeV is responsible for the 
acceleration limit. We find that this spectrum is in agreement with the 
EGRET observations (Thompson et.al. 1999) in the GeV bands. 
In the MeV bands, however, it is inconsistent with the 
observations. This is because the value of $\Gamma_{sat}$ makes 
the curvature 
spectrum with $E^2F\propto E^{\alpha}$,$\alpha\sim4/3$
 in the MeV bands, although the observations have $\alpha\sim1/3$.

In \S3, we have pointed out that if $W\ll\varpi_{lc}$, 
the curvature radiation from 
the outside of the gap is important.  
In the outside of the gap where $E_{||}$ is vanished, the particles lose their
 energy by the radiation, the spectrum of which  
 extends to the MeV band (dotted-line in Fig.2(a)). 
By inclusion of this emission, the total spectrum (solid-line in Fig.2(a)) is 
in good agreement with the EGRET observations. 

The calculated $total$ $\gamma$-ray spectrum for B1706-44 is shown in 
Fig.2 (b). We find that the calculated peak 
energy becomes slightly less than Vela 
because $E_{||}^{B1706}<E_{||}^{Vela}$ (\S\S4.1), and this peak energy also  
explains the observations.   
As mentioned in \S\S4.1, since the calculated $E_{||}$ becomes large 
as we assume the nearer distance to the pulsar,
the spectrum becomes hard if we adopt a nearer distance. 
For 1.8 kpc, the calculated spectrum appears 
to be consistent with the observations.
 However, we must assume a large cross-section area of the gap,
 $A_{\perp}=(10W)^2\sim(\varpi_{lc})^2$, to obtain the observed fluxes. 
Because the gap locates at $s\sim0.5\varpi_{lc}$ (Fig.1), 
such $A_{\perp}$ should be unrealistic.
 
Fig.2 (b) also shows the calculated IC spectrum from the gap. 
The sharp-cut off in the spectrum which corresponds to  
the acceleration limit appears around 10TeV. 
Since IR flux might be overestimated,
 TeV flux would  be less than $\sim10^{-12}\mathrm{erg\ cm^{-2}\
 s^{-1}}$. Moreover, 
this calculated flux hardly depends on the distance to the pulsar.
On these ground, we conclude that 
it is difficult to explain 
the unidentified TeV components observed by CANGAROO group 
(Kushida et.al., 2002) in the direction of B1706-44 with this model.
\section{Discussion}
In summary, we obtained the spectrum in good agreement with the EGRET 
observations for Vela by inclusion of the curvature radiation 
from the outside of the gap. We found that the observed peak energies 
for Vela and B1706-44 may imply that the almost the 
same currents in units of GJ value are running 
through the gap for both pulsars.   
 
From Fig.2 (a), we recognize that the calculated spectrum is 
inconsistent with the COMPTEL observations. If we try to 
explain this observations, we need the very small curvature radius 
as compared with the dipole, which is unlikely.
  Therefore, this MeV emission will be obtained by 
inclusion of the synchrotron emission by pairs.

In \S\S4.2, we showed that we need very large $A_{\perp}$ to explain the 
observations of B1706-44. 
This may be due to the small $j_{tot}$, 
about 20\% of the GJ current. 
However, if we adopt the nearly
 ($j_{tot}\sim1$) or super ($j_{tot}>1$) GJ current with this model, 
the gap width will be quenched.  

Quite recently, Hirotani et.al. (2002) showed that 
the value of the $\Gamma_{sat}$ 
given by eq.(\ref{lorent}) and also $\gamma$-ray fluxes
 calculated with this $\Gamma_{sat}$ are overestimated, 
 because particles in the gap do not immediately saturate. 
But, the general features of the gap model 
and the result that we must take account of
 the radiation from the outside of the gap to explain 
the observations above 100MeV are not altered.

$Acknowledgments$\quad We thank Drs. Tanimori and Kushida for useful 
discussion for B1706-44 and the CANGAROO group for 
holding the symposium. 
\section*{References}
\begin{verse}
 Cheng K.S., Ho C., Ruderman M.A.\ 1986, ApJ 300, 500

 Gotthelf E.V., Halpern J.P., Dodson R.\ 2002, ApJ 567, L125

 Kushida J. et.al.\ 2002, "The Universe Viewed in Gamma-Rays (Universal Academy Press)"
 
 Hirotani K., Shibata S.\ 1999, MNRAS 308, 54 (HS)

Hirotani K., Harding A.K., Shibata S. \ 2002,  to be submitted 

\"{O}gelman H., Finley J.P., Zimmermann H.U.\ 1993, Nat 361, 136

 Thompston D.J. et.al.\ 1999, ApJ 516, 297
\end{verse}
\end{document}